\documentclass[twocolumn,showpacs,superscriptaddress,preprintnumbers,amsmath,amssymb]{revtex4}
\usepackage{amssymb}
\usepackage{amsmath}
\usepackage{cases}
\usepackage{graphicx}
\usepackage{dcolumn}
\usepackage{bm}
\begin{document}
\title{First-principles study of temperature-dependent diffusion coefficients for helium in $\alpha$-Ti}
\author{Yong Lu}
\affiliation{LCP, Institute of Applied Physics and Computational Mathematics, Beijing
100088, People's Republic of China}
\author{Fawei Zheng}
\affiliation{LCP, Institute of Applied Physics and Computational Mathematics, Beijing
100088, People's Republic of China}
\author{Ping Zhang}
\thanks{Author to whom correspondence should be addressed. E-mail: zhang\_ping@iapcm.ac.cn}
\affiliation{LCP, Institute of Applied Physics and Computational Mathematics, Beijing
100088, People's Republic of China}
\affiliation{Beijing Computational Science Research Center, Beijing 100084, People's
Republic of China}

\begin{abstract}
The temperature-dependent diffusion coefficients of interstitial
helium atom in $\alpha$-Ti are predicted using the transition state
theory. The microscopic parameters in the pre-factor and activation
energy of the impurity diffusion coefficients are obtained from
first-principles total energy and phonon calculations including the
full coupling between the vibrational modes of the diffusing atom
and the host lattice. The climbing image nudged
elastic band (CINEB) method is used to search for the minimum energy
pathways and associated saddle point structures. It is demonstrated that the diffusion
coefficients within the \emph{xy} plane ($D_{xy}$) is always higher
than that along the \emph{z} axis ($D_{z}$), showing remarkable anisotropy. 
Also, it is found that the formation of helium dimer centered at the
octahedral site reduces the total energy and confines the diffusion of helium atoms.

\end{abstract}

 \maketitle
\section{Introduction}
Helium atoms could be produced in metals either through nuclear
reactions of energetic particles such as reactor neutrons and light
ions or via the radioactive decay of tritium \cite{Sciani1983}. Due
to the closed-shell electron configuration, helium is
inactive to metals and has extremely low solubility in metals.
Helium atoms diffuse rapidly in the interstitial region until it
reaches a trapping site. As the helium concentration increases by
direct implantation or neutron transmutation, they tend to
agglomerate into helium bubbles when several helium atoms migrate to
the same trapping site, which may substantially deteriorate the
mechanical properties of metals. Particularly, at high homologous
temperatures, helium bubbles can cause intergranular embrittlement,
cave, and swelling \cite{Schroeder1981,Trinkaus2003}. Many experimental and analytical investigations of helium bubble growth
and helium diffusion in metals have been presented \cite{Sciani1983,
Thomas1979, Schroeder1981,
Trinkaus2003,Vassen,Singh1992,Rajaraman1994,Glam2009}. Although the
experiments can give the formation energy of helium bubbles and
overall diffusion coefficients, generally they cannot determine the
microscopic physical processes involved in the formation and
diffusion steps, which are quite important for basic understanding
and practical applications. First-principles calculations and
molecular dynamic simulations, which have been widely used in the
study of solid-state diffusion, can help to track the microscopic
diffusion processes and give rise to specific quantitative values
involved in the atomistic processes, where the experimental
measurements cannot reach
\cite{Wilson1981,Wimmer,Mantina1,Chen,Mantina2,Zhang}.

As an important material to store and retrieve hydrogen, titanium
has many important applications in aircraft construction and
aerospace engineering, as well as in the chemical industry. While,
there are considerable restrictions in its applications
under hydrogen-containing environments. Helium atoms are usually
introduced into titanium through tritium decay, then titanium
containing tritium becomes doped with $^{3}$He, which tends to
agglomerate into bubbles resulting in a severe deterioration in the
mechanical properties. At room temperature, titanium exhibits the
hexagonal $\alpha$ phase, whereas, it undergoes phase transition
from $\alpha$-phase to $\beta$-phase (bcc structure) at the
temperature around 1155 K. This relatively low transition
temperature for Ti makes diffusion experiments difficult to be
solely carried out within the $\alpha$-phase. Furthermore, due to the
insolubility of helium, the knowledge about its diffusion behavior
in metals are very few in contrast to other light elements such
as H, C, N or O. To date, the experimental diffusion coefficients of
helium in $\alpha$ titanium are rather scarce \cite{Vassen}, and to
our knowledge, systematic \textit{ab initio} studies of hydrogen
diffusion in $\alpha$ Ti is still lacking in the literature. The
purpose of the present work is to comprehensively investigate the
atomic diffusion mechanism of helium in $\alpha$-Ti, and to obtain
the specific values of the corresponding energy barriers and
diffusion coefficients from first principles.

\section{Theory and methods}

In the most general form, the diffusion coefficient is expressed as
\begin{equation}
D=D_{0}e^{-Q/kT},
\end{equation}
where $Q$ is the activation energy, $D_{0}$ is a pre-factor, and $k$
is the Boltzmann constant. Following the Wert and Zener \cite{Wert},
the diffusion coefficient can be written as
\begin{equation}
D=n\beta d^{2} \Gamma,
\end{equation}
where $n$ is the number of nearest-neighbor stable sites for the
diffusing interstitial atom, $\beta$ is the jump probability in the
direction of diffusion, $d$ is the length of the jump projected onto
the direction of diffusion, and $\Gamma$ is the jump rate between
adjacent sites of the diffusing atom. According to the transition
state theory (TST) \cite{Eyring,Vineyard}, the jump rate is written
as
\begin{equation}
\Gamma=\frac{kT}{h}\frac{\prod^{3N-6}_{i=1}\left[1-exp(-h\nu^{0}_{i}/kT)\right]}{\prod^{3N-7}_{i=1}\left[\left[1-exp(-h\nu^{*}_{i}/kT)\right]\right]}e^{-\Delta H_{m}/kT},%
\end{equation}
where $\nu^{*}_{i}$ and $\nu^{0}_{i}$ are the vibrational
frequencies at the transition state and the ground state,
respectively. For the high temperature range ($h\nu_{i}/kT \ll  1$),
 all vibrational degrees of freedom approach a classical behavior and
the jump rate reduces to,
\begin{equation}
\Gamma=\frac{\prod^{3N-6}_{i=1}\nu^{0}_{i}}{\prod^{3N-7}_{i=1}\nu^{*}_{i}}e^{-\Delta H_{m}/kT}.%
\end{equation}

For the low temperature range ($h\nu_{i}/kT \gg  1$), the jump rate
can be expressed as \cite{Wimmer}
\begin{equation}
\Gamma=\frac{kT}{h}e^{-(\Delta H_{m}+\Delta E_{zp})/kT},%
\end{equation}
where $E_{zp}$ is the difference in zero-point energies between
ground state and transition state.

Using the phonon free energy at different states
\begin{equation}
F_{vib}=kT\int^{\infty}_{0}g(\nu)ln\left[2sinh\left(\frac{h\nu}{2kT}\right)\right],%
\end{equation}
the jump rate can be simply expressed as
\begin{equation}
\Gamma=\frac{kT}{h}e^{-\Delta F_{vib}/kT}e^{-\Delta H_{m}/kT},%
\end{equation}
where the zero-point energy is included in the $F_{vib}$ term.

The density functional theory (DFT) calculations are carried out
using the Vienna \emph{ab-initio} simulation package (VASP)
\cite{Kresse, Kresse2} with the projector-augmented-wave (PAW)
potential method \cite{Blochl}. The exchange and correlation effects
are described by the local density approximation (LDA). The cutoff
energy for the plane-wave basis set is 450 eV. We employ a
$3\times3\times2$ $\alpha$-Ti supercell containing 36 host atoms to
simulate helium migration in the $\alpha$-Ti matrix. To check the
convergence of the formation energies, we have also considered a
$4\times4\times3$ $\alpha$-Ti supercell containing 96 host atoms and
one helium atom. The integration over the Brillouin zone is carried
out on $5\times5\times5$ $k$-point meshes generated using the
Monkhorst-Pack \cite{Monkhorst} method, which proves to be
sufficient for energy convergence of less than 1.0$\times$10$^{-4}$
eV per atom. During the supercell calculations, the shape and size
of the supercell are fixed while all the ions are free to relax
until the forces on them are less than 0.01 eV {\AA }$^{-1}$. In
order to calculate the migration energies of helium in titanium,
each saddle-point structure and associated minimum energy pathway
(MEP) were calculated by employing the image nudged elastic band
(CINEB) method \cite{CINEB}. The vibrational frequencies were
computed using the density functional perturbation theory with the
help of PHONOPY package \cite{Togo}.

\section{Solution of helium atom}

\begin{figure}[ptb]
\includegraphics[width=0.3\textwidth]{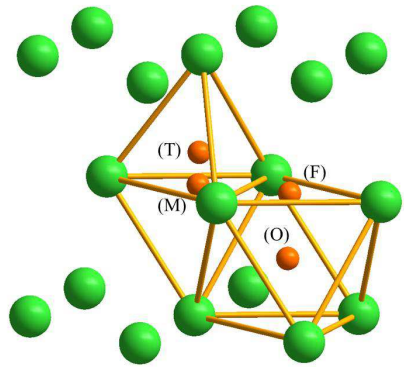}\caption{(Color online)
Visual structure of $\alpha$-Ti matrix marked with tetrahedral
(T) site, octahedral (O) site, the center of equilateral trigonal
face of octahedron (F) site, and the center of equilateral trigonal
face of tetrahedral (M) site. Large green and small
red balls represent for titanium atoms and interstitial sites for helium atom, respectively.}%
\label{fig1}%
\end{figure}

\begin{figure}[ptb]
\includegraphics[width=0.4\textwidth]{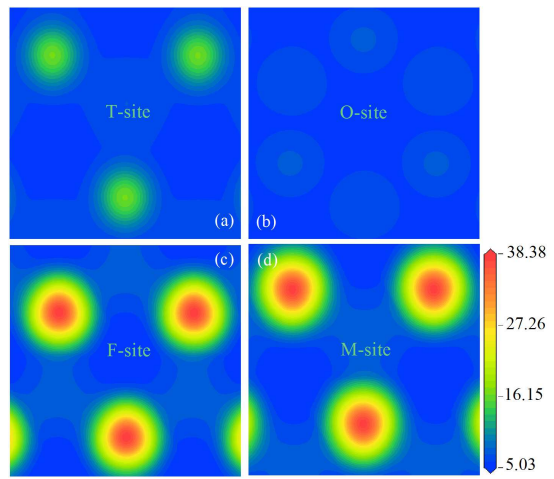}\caption{(Color online) Contour
plots of charge density distributions of pure Ti matrix inside the
(110) plane
crossing T-site (a), O-site (b), F-site (c), and M-site (d), respectively.}%
\label{fig2}%
\end{figure}

\begin{figure}[ptb]
\includegraphics[width=0.5\textwidth]{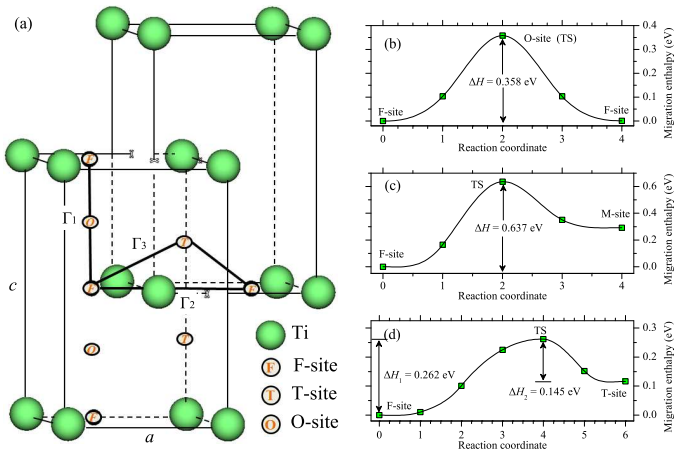}\caption{(Color online) Visual structure of
three different jump paths characterized by their jump rates
$\Gamma_1$, $\Gamma_2$, and $\Gamma_3$ for the He (a), and the
reference energy profiles for the He to diffuse along (b)
F$\rightarrow$O$\rightarrow$F, (c) F$\rightarrow$M, and (d)
F$\rightarrow$T paths.}%
\label{fig3}%
\end{figure}

\begin{table}[ptb]
\caption{Calculated formation energy ($E_{f}$) for helium atom
in various solution sites in hcp-type Ti matrix.}%
\begin{ruledtabular}
\begin{tabular}{cccccccccccccccc}
&$E_f^{T}$&$E_f^{O}$&$E_f^{F}$&$E_f^{M}$\\
\hline
36-atom cell&2.501&2.746&2.385&2.677\\
96-atom cell&2.499&2.747&2.382&2.675\\
\end{tabular} \label{a}
\end{ruledtabular}
\end{table}

The formation energy of interstitial helium atom in titanium matrix
is calculated by
\begin{equation}
E_{f}({\rm He})=E({\rm Ti},n{\rm He}) -n\mu -E({\rm Ti}),%
\label{eq7}
\end{equation}
where $E({\rm Ti},n{\rm He})$ is the total energy of the supercell
containing $n$ interstitial helium atoms, $\mu$ is the chemical
potential of helium atom, $E({\rm Ti})$ is the energy of the Ti
supercell without interstitial atoms.

The favorable solution sites for helium atom in metals are those
positions of high-symmetry or low charge density. From an
intuitional consensus, the octahedral site (O-site) with six nearest
Ti atoms and the tetrahedral site (T-site) with four nearest Ti
atoms are strong candidates, as marked in Fig. 1. In addition to
O-site and T-site, there are two other interstitial sites in the
hcp-type Ti matrix, i.e., the center of equilateral trigonal face
shared by two adjacent O sites (F-site) and  the center of
equilateral trigonal face shared by two adjacent T sites along $z$
axis (M-site), as depicted in Fig. 1. We calculated the formation
energies of one helium atom in these four types of solution site
using the 36-atom model and 96-atom model respectively, and the
results are organized in Table I. Clearly, these two models show the
similar convergence in energy, implying that the 36-atom supercell
is sufficient to assume the simulation task.

The positive formation energies indicate that the dissolution of He
in Ti is endothermic. The formation energy of helium in titanium
matrix (ranging from 2.385 eV to 2.746 eV) is much lower than that
in bcc-type Fe (ranging from 4.22 eV to 5.36 eV) and fcc-type Ni
metals (around 4.50 eV) \cite{Wilson,Willaime,Seletskaia}. The
favorable site in energy for helium atom in titanium matrix is the
F-site. Unexpectedly, the O-site, supplying the maximum space for
helium atom among the four interstitial sites, has the highest
formation energy. This behavior of helium is different from that of
H atom in titanium, where O-site is more favorable in energy
\cite{Lu}. To a great extent, the electronic distribution and the
interaction between He atom and its surrounding Ti atoms make main
contribution in formation energy. Thus, we first plot the charge
density of (001) planes crossing the four different sites for pure
Ti matrix, as shown in Fig. 2. By comparing the four contour plots
of charge density [Figs. 2(a)-(d)], the position with the lowest
charge density is the F-site, and the M-site occupies the highest
charge density. The charge density in O-site is somewhat higher than
that in F-site, which can be further confirmed by comparing the
reference isosurface (not shown here). From the viewpoint of
electron distribution, F-site is the most favorable site for helium
atom, since the interstitial light elements tend to locate at the
site with lowest electron density \cite{Norskov}. Furthermore, we
calculated the interaction energy between interstitial helium atom
and its surrounding host atoms according to the following
expression
\begin{equation}
E_{int}({\rm He - Ti})=E({\rm Ti,He})-E({\rm Ti})^{*}-E({\rm He}),%
\label{eq7}
\end{equation}
where $E({\rm Ti,He})$ is the total energy of optimized Ti supercell
containing one He atom, $E({\rm Ti})^{*}$ is the total energy of
the supercell containing only Ti atoms with the same atom position
and cell parameters as those of $E({\rm Ti, He})$, and $E({\rm He})$
is the energy of one isolate helium atom. The reference results are
organized in Table II. Clearly, the interaction energy $E_{int}$ in
F-site is the lowest, while the highest energy appears at the
O-site. By comparing the energy, the F-site is still the most
favorable site for helium atom. The interaction between interstitial
helium atom and host atoms will cause the deformation of Ti lattice,
further increasing the formation energy. The deformation energy can
be evaluated by
\begin{equation}
E_{def}({\rm Ti})=E({\rm Ti})^{*}-E({\rm Ti}),%
\label{eq7}
\end{equation}
and the results are also listed in Table II. Due to the strong
interaction between He and Ti atoms, the O-site has the largest
deformation energy. The F-site and T-site have low deformation
energy, coinciding with the relatively weak interaction between He
and host atoms in these two sites.

\begin{table}[ptb]
\caption{Interaction energy ($E_{int}$) and deformation
energy ($E_{def}$) for helium atom in different solution sites.}%
\begin{ruledtabular}
\begin{tabular}{cccccccccccccccc}
&F-site&O-site&T-site&M-site\\
\hline
$E_{int}$(He-Ti)&1.952&2.169&2.074&2.130\\
$E_{def}$(Ti)&0.432&0.578&0.427&0.547\\
\end{tabular} \label{a}
\end{ruledtabular}
\end{table}

\section{Diffusion in titanium}
\begin{table*}[ptb]
\caption{Calculated electronic energy $E_{el}$ and zero-point energy
$E_{zp}$ for interstitial helium atom in 36-Ti matrix. All energies
are in kJ/(mol of Ti$_{36}$(He)). The electronic energy of the
elements in their standard states, i.e., Ti$_{36}$, is taken as reference.}%
\begin{ruledtabular}
\begin{tabular}{cccccccccccccccc}
&Ti$_{36}$&Ti$_{36}$He(F)&Ti$_{36}$He(T)&Ti$_{36}$He(O)&Ti$_{36}$He(M)&Ti$_{36}$He(TS$_{\mathrm{FT}}$)&Ti$_{36}$He(TS$_{\mathrm{FM}}$)\\
\hline
$E_{el}$&0&229.63&240.80&264.49&257.78&254.87&290.99\\
$E_{zp}$&119.39&124.98&123.11&113.59&121.82&122.73&121.86\\
\end{tabular} \label{a}
\end{ruledtabular}
\end{table*}

\begin{figure}[ptb]
\includegraphics[width=0.4\textwidth]{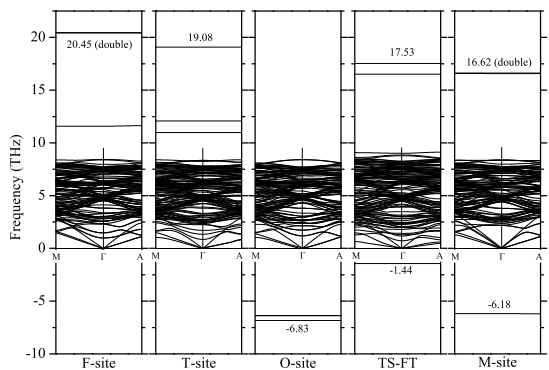}\caption{Calculated phonon dispersions of a 36-atom Ti supercell with helium
atom in the F-site, T-site, O-site, M-site and the transition states
along the F$\rightarrow$T (TS-FT) and F$\rightarrow$M (TS-FM) paths,
respectively. The modes with imaginary frequencies at the transition
states correspond to the motion of the helium atom across the
barriers.}%
\label{fig4}%
\end{figure}

\begin{figure}[ptb]
\includegraphics[width=0.3\textwidth]{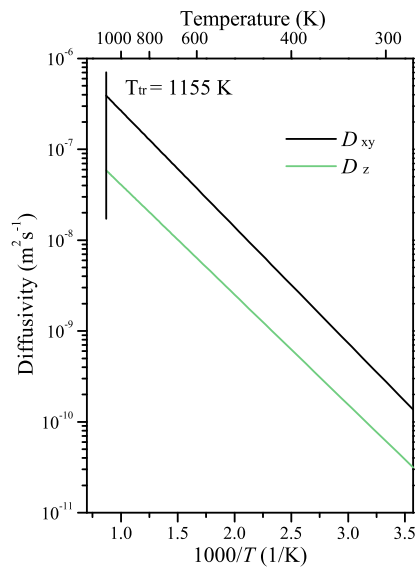}\caption{(Color online)
Calculated diffusion coefficients of atomic helium in $\alpha$-Ti.
The T$_{tr}$=1155 K stands for the phase transition temperature of
metal Ti from $\alpha$-phase to $\beta$-phase.}%
\label{fig5}%
\end{figure}

\begin{table*}[ptbptb]
\caption{Calculated diffusion pre-factors ($D_{0}^{xy}$ and
$D_{0}^{z}$) and activation energy ($Q_{xy}$ and $Q_{z}$) for helium
atom in $\alpha$-Ti. The temperatures represent the ranges over
which diffusion coefficients were fit to extract corresponding
pre-factors and activation energies. For comparison, experimental
and molecular simulation results are also listed.}%
\begin{ruledtabular}
\begin{tabular}{cccccccccccccccc}
Method&$D_{0}^{xy}$ (m$^{2}$/s)&$D_{0}^{z}$ (m$^{2}$/s)&$Q_{xy}$ (kJ/mol)&$Q_{z}$ (kJ/mol)&T (K)\\
\hline
This study&5.06$\times$10$^{-6}$&6.63$\times$10$^{-7}$&24.48&23.16&280-850\\
MD$^{a}$&2.1$\times$10$^{-8}$&2.6$\times$10$^{-7}$&17.34&17.34&298-667\\
\end{tabular} \label{a}
$^{a}$ Reference \cite{Chen}
\end{ruledtabular}
\end{table*}

From the calculation of formation energy above, we have confirmed
that the most favorable site for helium atom in $\alpha$-Ti matrix is the
F-site. In the following, we will investigate the diffusion paths
for interstitial helium atom.

As illustrated in Fig. 3 (a), we set the F-site as initial site and
choose three candidate jump types characterized by their
corresponding jump rates $\Gamma_1$, $\Gamma_2$, and $\Gamma_3$. A
diffusive jump of $\Gamma_1$ connects two nearest-neighbor F sites
in $z$-axis direction. $\Gamma_2$ and $\Gamma_3$ are two jump types
of helium atom within the basal $xy$ plane, where $\Gamma_2$
connects these two nearest-neighbor F sites directly and $\Gamma_3$
via the T site. We shall determine the more favorable jump type
between $\Gamma_2$ and $\Gamma_3$ by comparing their migration
enthalpy. To estimate the migration enthalpy for helium atom
diffusion between interstitial sites, the three paths are evenly
divided into several segments. We placed He atom at the appropriate
position every equal distance. By employing the CINEB method, the
activation energy profiles vs diffusion distance are obtained and
shown in Figs. 3(b)-(d). For the diffusion path $\Gamma_1$, the
energy is found to display a single maximum, corresponding to a
saddle point at the high-symmetry position located half way between
neighboring F sites, i.e., the O site acts as a transition state
role. The energy profile can be well described by a sinusoidal curve
with an migration enthalpy of 0.358 eV. For jump $\Gamma_2$, we find
that the middle point of diffusion path coincides with M site when
the relaxation process completed. Actually, the jump $\Gamma_2$ thus
connects two nearest-neighbor F sites via the M site. Due to the
symmetry of the diffusion curves, we plot here only the ones of
F$\rightarrow$M for $\Gamma_2$ path and F$\rightarrow$T for
$\Gamma_3$ path respectively, as shown in Figs. 3(c) and 3(d).
Evidently, the migration enthalpy of the $\Gamma_3$ path (0.262 eV)
is much lower than that of $\Gamma_2$ (0.637 eV). That is, the
helium atom tends to diffuse between F sites through the $\Gamma_3$
path within the basal $xy$ plane.

The lattice vibrational energy also plays a great role in the
migration energy, especially in the high temperature range. In the
following, we will calculate the phonon dispersions of these
different states using the density functional perturbational theory,
and obtain the reference lattice vibrational energy. The
3$\times$3$\times$2 Ti supercell matrix is employed and one helium
impurity is placed at the F-site, T-site, O-site, M-site, and the
transition states of F$\rightarrow$T (TS-FT), respectively. The
phonon dispersions of all these atomic structures including the pure
36-Ti matrix are calculated and shown in Fig. 4. Obviously, when a
helium atom occupies the F site, the He-related phonon dispersions
have the highest frequencies, which show a double degenerate
dispersionless branch at 20.4 THz. The F site locates at the
geometric center of triangle formed by its three nearest Ti atoms,
thus the interstitial space is much more confined with respect to
other states. At the T site, the highest He-related frequency
decreases to 19.1 THz and the degeneracy vanishes. In both cases, the
frequencies are positive, indicating a true minimum in the energy.
By definition, the transition state is characterized by the
occurrence of one negative eigenvalue in the dynamical matrix,
showing the imaginary frequency in the phonon dispersion, which
corresponds to a motion of helium atom along the diffusion path.
When helium locates at the transition state along the
F$\rightarrow$T path, we note that a Ti-related phonon branch
between 8.5 and 9.0 THz is obviously off from the bulk of Ti-related
phonon branches. This branch is related to motions of Ti atoms,
which couple with motions of the He impurity. At the octahedral
site, the He-related branches are all negative, locating at the $-$6.8
THz nearby. The O-site is the point of maximum solution energy and
the helium atom is unstable at this point, tending to diffuse away
from it. From the phonon dispersions, we can evaluate the zero-point
energy, the temperature dependent enthalpy, and free energy. The
electronic energy and zero-point energy of these states are listed
in Table III. The zero-point energy at F site is 124.98 kJ/mol,
which is the highest one among all the considered states due to the
highest He-related branch. With the correction of zero-point energy,
the most favorable interstitial site in energy for helium is still
the F site at ground state. Finally, we can confirm that the helium
atom diffuse in Ti matrix by $\Gamma_1$ jump path along the $z$ axis
and $\Gamma_3$ jump path within the $xy$ plane, respectively.

\section{Diffusion coefficients}
Under the assumption of the above jump paths and according to Eq. (2), the
two diffusion coefficients in the $hcp$-structured $\alpha$-Ti can
be expressed as
\begin{equation}
D_{xy}=a^{2} \Gamma_{1}%
\end{equation}
and
\begin{equation}
D_{z}=(\frac{c}{2})^{2} \Gamma_{3},%
\end{equation}
respectively, where $a$ and $c$ are the lattice parameters of $\alpha$-Ti.
Diffusion coefficient presents more quantitative description on the
diffusion features. In Fig. 5 we show the Arrhenius plot, i.e.,
$\ln(D)=\ln(D_0)-Q/kT$, of the computed diffusion coefficients
$D_{xy}$ and $D_z$. As shown in Fig. 5, the diffusion of helium in
$\alpha$-Ti matrix shows remarkably linear behavior and thus we can
obtain the activation energy and prefactor by linear fitting. In
Table IV we listed our fitting results of pre-factors and activation
energies.

A linear fit between 280 K and 850 K of the calculated data gives
$Q_{xy}$=24.48 kJ/mol and $Q_{z}$=23.16 kJ/mol, both somewhat higher
than the molecular dynamic simulation results of 17.34 kJ/mol
\cite{Chen} obtained using the Einstein relation \cite{Boisvert}.
The diffusion coefficients of these two directions are very
anisotropic, corresponding to the pre-factors
of $D_{0}^{xy}$=5.06 $\times$10$^{-6}$ m$^{2}$/s and $D_{0}^{z}%
$=6.63$\times$10$^{-7}$ m$^{2}$/s, respectively. This is consistent
with the qualitative observation of helium migration enthalpy, where
the helium atom is easier to overcome the migration enthalpy
within the $xy$ plane (cf. Fig. 3). Note that the theoretical
diffusion coefficients obtained here should be the upper bound of
helium migration in $\alpha$-Ti matrix. Due to the existence of
vacancies or dislocations in a real crystal, as well as the
formation of helium bubble, the diffusion coefficients will be
reduced to some extent. To this end, we also made a relevant test to
elaborate the effects of clustering behavior of two helium. We set
two helium atoms at the adjacent F sites along the $z$ axis, and
then perform the relaxation of the system until the largest force on
any atom is smaller than 0.01 eV/{\AA}. We find that two helium
atoms form a dimer along the $\langle$001$\rangle$ direction
centered at the octahedral site. The formation energy is just 2.05
eV per He atom, lower than that of one helium case (cf. Table I).
The formation of helium dimer will confine the diffusion of helium
atom in Ti matrix, reducing the diffusion coefficients.

\section{Conclusion}
In summary, we have systematically studied the temperature-dependent
diffusion coefficients of helium atom in $\alpha$-Ti using
transition state theory with accurate first-principles total energy
and phonon calculations. It is found that the most stable solution
site for He in Ti matrix is the F site, with the lowest formation
energy of 2.385 eV. Two minimum energy pathways and associated
saddle point structures are determined by using the CINEB method. Within the
$xy$ plane, the helium atom diffuse between adjacent solution sites
by crossing the T meta-stable site, while along the $z$ axis it
should pass the transition state of O site. The obtained diffusion
coefficients within the \emph{xy} plane ($D_{xy}$) and along the
\emph{z} axis ($D_{z}$) show remarkable anisotropy, and the helium
atom is more easily to diffuse within the $xy$ plane. The formation
of helium dimer centered at the octahedral site reduces the total energy compared to the case of two isolate helium atoms, which confines
the diffusion of helium and further reduce the diffusion
coefficients. Without considering the vacancies or helium bubble
effects, our theoretical diffusion coefficients should be
an upper bound on the helium diffusion in Ti matrix, which provides
a good reference for future experimental measurements on
$\alpha$-Ti.

\begin{acknowledgments}
This work was supported by NSFC under Grant No. 51071032 and by CAEP
Foundations for Development of Science and Technology under Grant
No. 2011A0301016.
\end{acknowledgments}

\end{document}